\documentclass[twocolumn,superscriptaddress,showpacs,preprintnumbers,ams,symb,prl]{revtex4}

\usepackage{graphicx}% Include figure files
\usepackage{dcolumn}% Align table columns on decimal point
\usepackage{bm}% bold math
\usepackage{amsmath}
\usepackage{hyperref}
\usepackage{color}
\usepackage{SIunits}

\newcommand{\caco}{$\mathrm{Ca_3Co_2O_6 }$}

\begin{document}

\title{Incommensurate magnetic ground state revealed by RXS in the frustrated
spin system \caco}

\author{S. Agrestini}
\affiliation{Department of Physics, University of Warwick, Coventry, CV4 7AL, UK}

\author{C. Mazzoli}
\affiliation{European Synchrotron Radiation Facility, BP 220,
38043 Grenoble Cedex 9, France }

\author{A. Bombardi}
\affiliation{Diamond Light Source Ltd., Rutherford Appleton
Laboratory, Chilton-Didcot OX11-0QX, UK }

\author{M.R. Lees}
\affiliation{Department of Physics, University of Warwick, Coventry, CV4 7AL, UK}

\date{\today}

\begin{abstract}
We have performed a resonant x-ray scattering study at the Co pre-K
edge on a single crystal of \caco. The measurements reveal an abrupt
transition to a magnetically ordered state immediately below
$T_N$~=~ 25 K, with a magnetic correlation length in excess of 5500
\AA\ along the $c$-axis chains. There is no evidence for
modifications to the Co$^{3+}$ spin state. A temperature dependent
modulation in the magnetic order along the $c$-axis and an unusual
decrease in the magnetic correlation lengths on cooling are
observed. The results are compatible with the onset of a partially
disordered antiferromagnetic structure in \caco.

\end{abstract}


\maketitle

\caco\ provides us with an opportunity to study the phenomena of low-dimensional magnetism and topological magnetic frustration in a single compound. \caco\ consists of chains, made up of alternating face-sharing octahedral (CoI) and trigonal prismatic (CoII) CoO$_{6}$ polyhedra, running along the $c$-axis and arranged in a triangular lattice in the $ab$-plane \cite{Fjellvag96}. The different Co environments leave the Co$^{3+}$ ions on the CoI sites in a low-spin (S = 0) state, and those on the CoII sites in the high-spin (S = 2) state \cite{Sampa04,Burnus06}. Crystalline electric fields also lead to a very strong Ising-like anisotropy with the moments preferentially aligned along the $c$-axis \cite{Kageyama97,Maignan00}.
The magnetic exchange is ferromagnetic (FM) along the chains and antiferromagnetic (AF) in the buckled  ${ab}$-plane \cite{Aasland97,Maignan00} making this system a rare example where Ising ferromagnetic chains are coupled antiferromagnetically on a triangular lattice.

Specific heat and magnetization measurements show the onset of long range magnetic order at $T_N$~=~25~K \cite{Aasland97,Hardy03}. AF interactions within the \emph{ab}-plane lead to a geometrical frustration of the magnetic structure and many degenerate spin configurations are possible, giving rise to highly susceptible dynamical states \cite{Wannier50}.
For this reason, despite several neutron diffraction studies \cite{Aasland97,Kageyama98,Petrenko05}, the nature of the magnetic ground state is still not fully understood. An intriguing result of the neutron experiments is the observation of a decrease in the integrated intensity of the magnetic peaks on cooling. One model \cite{Kageyama97b} describes the system below $T_N$ as a ``partially disordered antiferromagnet" (PDA), where two-thirds of the FM chains are coupled antiferromagnetically while the remaining third are incoherent. Recent neutron diffraction measurements \cite{Agrestini07} have shown that the long range magnetic structure in \caco\ is PDA and not ferrimagnetic \cite{Aasland97} but that more complex models should be considered to fully describe the magnetic order.

Below $T_S$$\sim$ 8 K a frozen spin (FS) state appears, where the
application of a magnetic field parallel to the chains leads to
striking new features; the appearance of  hysteresis in the ${M(H)}$
curves together with a succession of magnetization steps with a
roughly constant field spacing \cite{Kageyama97,Maignan00,Hardy04}.
These irreversible steps are related to metastable states whose
dynamics strongly depend on both the thermal and magnetic history
\cite{Hardy04}. Magnetization data showed relaxation effects that
are probably related to the frustration and the slow dynamics
associated with spin reversals in the FM chains. The existence of
the steps in the ${M(H)}$ curves could be due to the nearly
degenerate energy of different arrangements of the Co magnetic
moments in the triangular plane perpendicular to the chains
\cite{Maignan00,Kudasov06}. An alternative mechanism
analogous to quantum tunneling of magnetization in molecular magnets
has been proposed, with a fragmentation of the FM chains into finite spin units \cite{Maignan04}.

In this letter we present the first resonant x-ray scattering (RXS)
study of \caco. The photon energy spectra measured on magnetic
reflections are temperature independent indicating that there is no
change in the Co electronic configuration below $T_N$ nor a
detectable change in the structural point-symmetry at the Co sites.
We observe a temperature dependent modulation in the FM order along
the $c$-axis. A counter intuitive decrease in the magnetic
correlation lengths is observed on cooling. These results,
compatible with both theoretical calculations for PDA systems and
neutron diffraction measurements, open the way to novel descriptions
of the magnetic structure of this complex system.

Single crystals of \caco\ were grown by a flux method. The same single crystal $5~\times~2~\times~1$~mm$^3$ with the largest face perpendicular to the $(1\,1\,0)$ direction was used for all the RXS experiments.
Its high-quality was confirmed by x-ray diffraction, EDX, magnetization and specific heat measurements.
The RXS experiments were performed at the magnetic scattering beamline ID20 \cite{Paolasini07} at the ESRF (Grenoble). The beamline optics were optimized at 7.7 keV close to the Co K-edge.

Experiments were performed by using the natural ($\sigma$) incident
synchrotron polarization, with the sample mounted in a displex
cryostat. The diffractometer was operated in the vertical plane
scattering mode with an azimuth set-up, to allow for a sample
rotation about the scattering vector. The integrated intensity of
the reflections was measured using a photon counting avalanche
photodiode detector. The polarization of the reflected beam was
linearly analyzed by rotating the scattering plane of a highly
oriented $\langle 00L \rangle$ pyrolitic graphite plate.

The lattice parameters (a=9.062 {\AA} and c=10.39 {\AA} at $T$~=~30
K) were found to be in good agreement with neutron data
\cite{Aasland97,Fjellvag96}. The charge reflections have a FWHM of
less than 0.03 degrees and show no changes in intensity, shape, or
position. In non-resonant conditions, no extra charge reflections
appear on cooling and no sign of a structural transition was
detected down to $T$~=~2 K. Below $T_N$~=~25 K, antiferromagnetic
reflections with $-h+k+l \neq 3n$ appear. These reflections were
also observed with the neutrons and are generated by the onset of
long range magnetic order \cite{Aasland97,Kageyama98,Petrenko05},
which reduces the magnetic space group symmetry from
\emph{R$\overline{3}$c} to \emph{P3}.

\begin{figure}
\includegraphics[angle=-90,width=0.90\columnwidth]{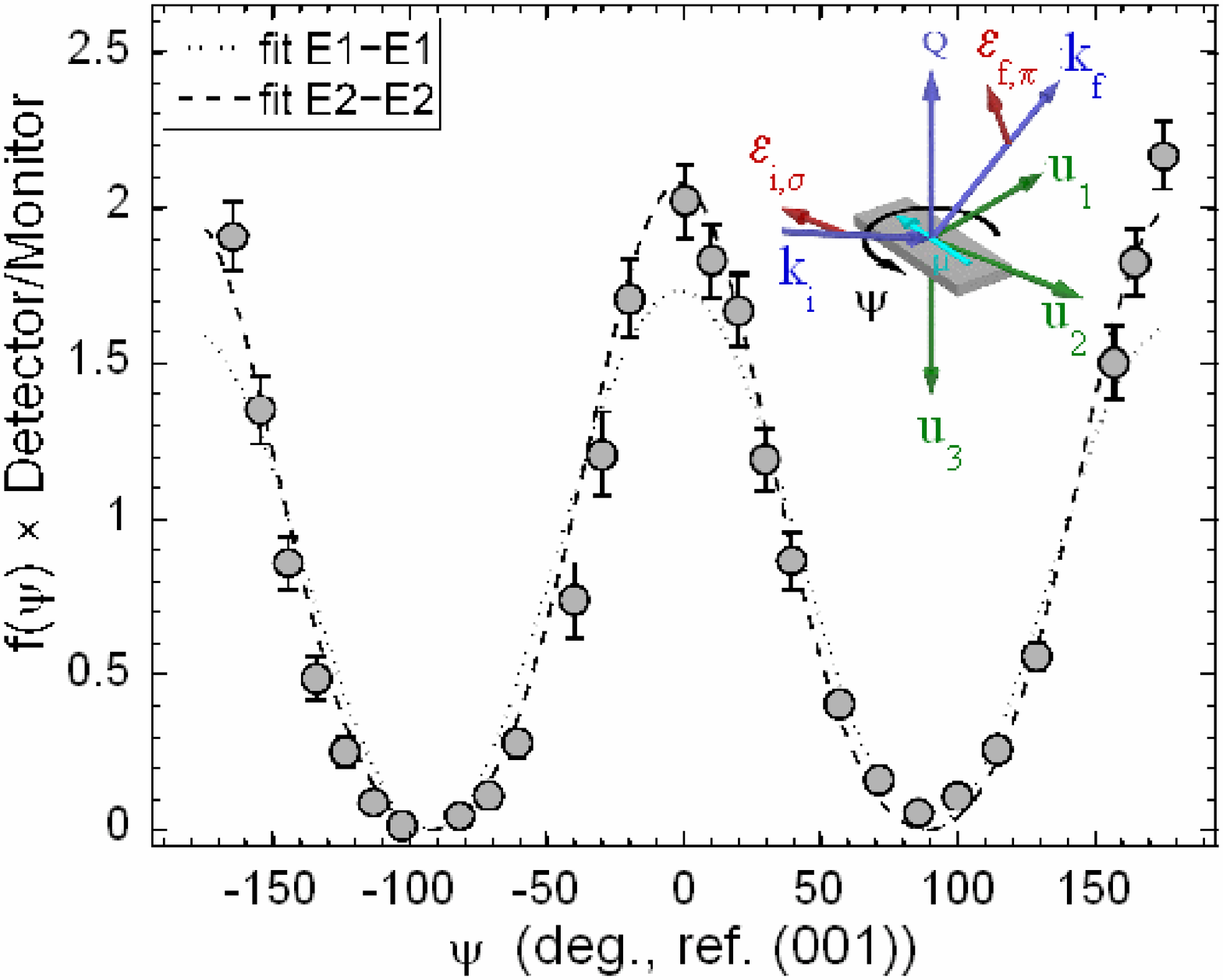}
\caption{\label{fig:azimuth} Azimuth scan on the $(3\,2\,0)$
reflection collected in the $\sigma-\pi$ channel at E=7.707 keV, $T$
= 4.2~K. Azimuth reference $(0\,0\,1)$. The dotted line shows a
purely E1-E1 fit and the dashed line is a fit including E1-E1 and
E2-E2 contributions. The inset shows the geometry of the scattering
and the fixed reference system $u_1,u_2,u_3$ used.}
\end{figure}

In the rest of this letter we focus on the $(3\,2\,0)$ reflection,
as all the magnetic reflections measured provide the same physical
information. Fig.~\ref{fig:azimuth} shows the azimuth scan in
resonant conditions (E=7.707 keV) collected in the rotated channel
with respect to the $(0\,0\,1)$ azimuthal reference direction at
$T$~=~4.2~K. The fits to the data show the elastic scattering
amplitude is dominated by electric dipolar transitions
\cite{Hill96,Carra94}, with higher order contributions coming from
quadrupolar electrical transitions. An analysis of the spherical
multipoles that can contribute to the scattering in the case of the
point group {\emph 3} is given in ref. \cite{Carra94}. Here the
reference system and the coordinates given in ref. \cite{Hill96} are
adopted. If just the dipolar contribution is considered, each
magnetic ion's contribution to the scattering is zero in the
unrotated $\sigma \sigma$ channel and is proportional to $\vec{k}
\cdot{} \hat{z}$ in the rotated channel $\sigma \pi$, where
$\vec{k}$ is the incident wavevector and $\hat{z}$ a vector
describing the magnetic moment. In the reference system sketched in
the inset of Fig.~\ref{fig:azimuth}, the magnetic structure factor
is proportional to $z_1 \cos(\psi) \cos ( \theta ) - z_3 \sin (
\theta )$, where $\hat{z}=(z_1,z_2,z_3)$ are the components of the
magnetic moment along {$u_1,u_2,u_3$}, $\theta$ is the Bragg angle
and $\psi$ the azimuth angle.

In our case the proportionality coefficients, embedding both the
magnitude of the magnetic moment and the matrix element between the
ground state and the intermediate state, are considered as free
parameters in the fit. Our data confirm the strong Ising character
of the system. With the moment aligned along the $c$-axis, $z_3$ is
zero and the expected azimuthal dependency of the $(3\,2\,0)$
reflection is given by the dotted line in Fig.~\ref{fig:azimuth}.
Allowing the magnetic moment to have an in-plane component does not
improve the fit and contradicts a number of other observations. A
significant improvement to the fit (dashed line in
Fig.~\ref{fig:azimuth}) is obtained by considering the quadrupolar
E2-E2 channel. In the limit of a zero $z_3$ component, two
quadrupolar higher order terms, proportional to
$z_1^3\cos^3(\psi)\cos^3(\theta)-z_1\cos(\psi)z_2^2\sin^2(\psi)\sin(2\theta)\sin(\theta)$
and to $z_1\cos(\psi)(z_1^2\cos^2(\psi)\cos(\theta)\sin^2(\theta) +
z_2^2\sin^2(\psi)\cos(2\theta)\cos(\theta))$, contribute to the
signal. The need to include higher rank tensors in the analysis is
not surprising as the electronic {\emph d} states are energetically
localized in the pre-edge region.

\begin{figure}
\includegraphics[angle=0,width=0.90\columnwidth]{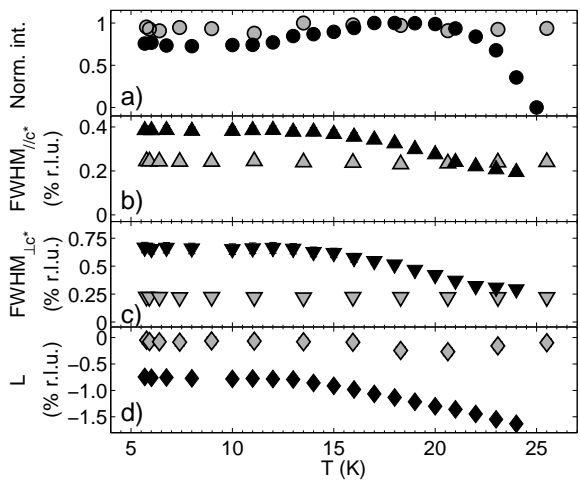}
\caption{\label{fig:Tdep} Evolution of the $(3\,2\,0)$ magnetic reflection (black symbols) versus $T$, taken at $\psi=0$ and E=7.707 keV. Panels contain a) the integrated intensity, b) the FWHM along $\vec{c}$ (L-scan), c) the FWHM in the $ab$-plane (HK-scan), and d) the evolution of the center position in the L direction. The $T$ dependence of the corresponding parameters for the $(3\,3\,0)$ charge peak (gray symbols) are shown for comparison.}
\end{figure}

The temperature dependence of several quantities characterizing the magnetic reflections measured at $\psi=0$ are reported in Fig.~\ref{fig:Tdep} (black symbols). As a comparison, the same quantities are presented for a charge reflection (gray symbols). The temperature dependence of the integrated intensity (Fig.~\ref{fig:Tdep}a) is unusual. It shows a broad maximum at about 18 K and decreases (by up to 25\% of the maximum value) on further cooling to 5 K confirming the presence of an anomalous reduction in the magnetic scattering intensity seen in neutron diffraction studies \cite{Aasland97,Kageyama98,Petrenko05}. However, whereas the neutron experiments measured a resolution limited magnetic peak at all the temperatures, the much higher reciprocal space resolution of the x-rays allows us to observe several new features of the magnetic order.

The FWHM of the $(3\,2\,0)$ reflection measured along the $c$-axis
and in the $ab$-plane are reported in Figs.~\ref{fig:Tdep}b and
~\ref{fig:Tdep}c respectively. The inverse of the FWHM is directly
related to the evolution of the magnetic correlation length and
allows us to directly probe the dimensionality of the magnetic
structure. The width of a magnetic reflection in an L-scan
[HK-scan], i.e. the intensity as a function of the wave-vector
offset along the $(0\,0\,l)$ [$(h\,k\,0)]$ direction from the Bragg
position, gives information on the magnetic correlation length along
the $c$-axis in the [$ab$-plane]. Immediately below $T_N$, both the
L- and the HK-scans produce magnetic peaks with a FWHM almost equal
to those obtained in corresponding scans of charge reflections,
which are in turn only slightly wider than the instrumental
resolution. This is a proof that the transition to a metastable long
range magnetically ordered state is very abrupt and that immediately
below $T_N$ there is a significant alignment of the magnetic moments
both in the $ab$-plane and along the $c$-axis direction. Lower
limits to the correlation lengths are $\simeq$~5500~\AA\ along the
$c$-axis and $\simeq$~2500~\AA\ in the $ab$-plane.

Usually the FWHM of magnetic reflections in the vicinity of the
N{\'e}el point are quite broad. On cooling, as the moment saturates
and the thermal motion of the spins is reduced, the system normally
becomes more correlated and the reflection widths decrease. Here we
observe an increase in the FWHM of the magnetic peaks on further
cooling, with the values reaching a maximum at 12 K and then
remaining almost constant down to base temperature. These data,
therefore, provide direct experimental evidence that on cooling
there is a reduction of the magnetic correlation length. Further
evidence for the complex nature of the magnetic ground state is
given by the results presented in Fig.~\ref{fig:Tdep}d. Clearly the
$(3\,2\,0)$ magnetic peak is not centered at L=0 but moves as the
temperature is changed. A similar change in the position with
temperature in the L-direction was observed for all the magnetic
peaks examined, with either a positive or a negative shift with
respect to the center of the Brillouin zone, whereas the H- and
K-scans revealed no incommensurate behavior. As shown in
Fig.~\ref{fig:Tdep}d, the lattice Bragg reflections do not change
their L positions with $T$.

The movement of the magnetic reflections is a characteristic of the
magnetic structure of the system, with the non integer L values
being a signature that the order along the $c$-axis is not purely
ferromagnetic. The modulation changes its periodicity in a
continuous way from 700 \AA\ at $T$~=~23~K to 1500 \AA\ at
$T$~=~5~K.  The incommensurate structure and the reduction in intensity cannot be due to a
modulation of the direction of the Co moment; the azimuth
dependence of the $(3\,2\,0)$ reflection at 20~K and at
6~K are identical. There must, therefore, be a modulation in the magnitude
and/or sign of the magnetic moment. A periodic repetition of low,
intermediate, and high spin states is quite improbable. More
realistic, is the hypothesis of a modulation of the magnetic
order-disorder in the FM spin chains.

Monte Carlo simulations
\cite{Wada83} have revealed that the PDA structure is actually a
metastable state: the disordering of one third of the FM spin-chains
supposed in the PDA structure, produces a frustration of the FM
intra-chain interaction. In other words, there is competition
between the AF inter-chain coupling, which prefers to
have one third of the chains in an incoherent state, and the FM
intra-chain coupling, which requires all the FM chains are ordered.
The effect is a continuous interchange of the roles among the chains
in the PDA structure as a function of time and space. Theoretical
calculations reveal that these fluctuations can generate a time
dependent randomly modulated PDA structure (RMP) \cite{Matsubara83}.
Our data do not show an RMP phase, but a modulation with a well
defined periodicity. One should note that the Monte Carlo
calculations do not include defects that might stabilize one
particular structure and were performed by considering a FM
intra-chain coupling of the same order of magnitude as the AF
nearest-neighbor inter-chain coupling. In \caco, the FM coupling is
much stronger than the AF coupling. We hypothesize that in this
material the modulation of the PDA structure is not random, but that
a periodic repetition is stabilized. The continuous evolution of the
propagation vector on cooling seems to indicate that the thermal
energy above 12~K allows \caco\ to explore different metastable
configurations characterized by a propagation vector closer and
closer to (0,0,0), whereas below 12~K, the system is trapped, at
least on the timescale of our observations, in one state.

\begin{figure}
\includegraphics[width=0.8\columnwidth]{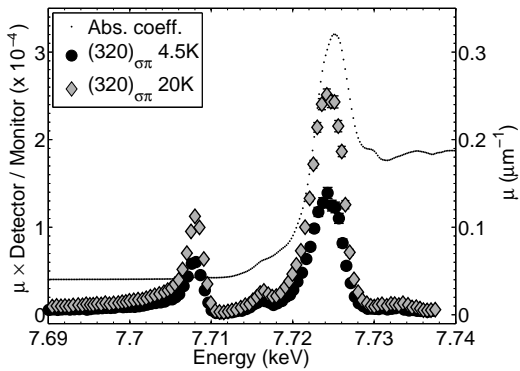}
\caption{\label{fig:spectra} Photon-energy dependence around the Co
K-absorption edge of the intensity of the $(3\,2\, 0)$ magnetic
reflection. Data were collected in the $\sigma-\pi$ channel at $T$ =
4.5 (black circles) and 20~K (gray diamonds). Corrections for
self-absorption have been applied. The absorption coefficient given by the fluorescence yield is shown
by the dotted line.}
\end{figure}

One possible origin of the decrease in the intensity of the magnetic
peaks is a reduction of the effective magnetic moment due to a
transition of the CoI ion from a high spin (S=2) to an intermediate
(S=1) spin state. Such a transition, however, should induce a change
in the Co density of states that would be reflected in the energy
spectra associated to the resonant magnetic reflection. To
investigate this possibility, the incident photon energy dependence
of the intensity of the $(3\,2\,0)$ magnetic reflection was measured
at $T$~=~4.5 and 20~K (Fig.~\ref{fig:spectra}). The spectra provide
a direct access to the p-d empty density of states of the CoII site
convoluted with the core-hole lifetime and the experimental
resolution. They both exhibit a narrow pre-edge feature and two
broader features between 7.71 and 7.73 keV. The main features of the
spectra are in good agreement with ref. \cite{Wu05} and provide
experimental evidence of the mixing of the p and the d states over a
broad energy range. Such a hybridization results from the absence of
the inversion symmetry at the CoII site, a condition not satisfied
at a perfectly octahedral site. Therefore a large dipolar
contribution is present over the whole density of states, even
though only the d states are magnetically active. The mainly dipolar
character of the transition together with the large orbital moment
measured experimentally \cite{Burnus06} account for the unusually
large enhancement observed over the non-resonant magnetic signal.
Apart from a scaling factor related to the intensity decrease at low
temperature the two spectra are indistinguishable ruling out a spin
state transition. No change in the number or type of magnetic
reflections observed, eliminates the possibility of a modification
to the long range magnetic structure with a transfer of intensity to
new reflections or different polarization channels.

The reduction in magnetic signal is due to a shift of intensity into
a diffuse scattering as seen in our neutron diffraction data
\cite{Agrestini07}. Our high resolution RXS measurements would not
detect such a broad signal. The presence of an increasing volume of
material with shorter range magnetic order (30 \AA) is consistent
with the picture of a metastable PDA structure presented above.

In conclusion, we report the first experimental evidence for the
modulated nature of the PDA order in \caco, a feature that until now
has only been predicted by theoretical calculations. Several
characteristics of the experimental data suggest that \caco\ becomes
more disordered as the temperature is reduced. The increased
disorder is not due to an electronic change occurring at the Co site
but is a co-operative phenomenon. At temperatures just below $T_N$,
a weaker inter-chain exchange allows extended units of FM aligned
magnetic moments to form along the $c$-axis. On cooling, the AF
in-plane coupling and the triangular geometry are incompatible with
the FM order within the chains. This produces smaller spin units
within the chains. The movement of the magnetic peak positions, as
the system searches for an energetically favorable configuration
slightly away from positions commensurate with the lattice, seems to
support this interpretation. A reduction in the $T$ dependence of
the various parameters characterizing the magnetic state below 12~K
reflects a slowdown in the evolution of the magnetic state as the
system nears the FS state.

This work was supported by a grant from the EPSRC, UK (EP/C000757/1). We acknowledge the European Synchrotron Radiation Facility for provision of beam time.

\end{document}